\documentclass[11pt]{article}
\usepackage{moriond,epsfig}

\def\Journal#1#2#3#4{{#1} {\bf #2}, #3 (#4)}

\def\PLB{{\em Phys. Lett.}  B}
\def\PRL{\em Phys. Rev. Lett.}
\def\PRC{{\em Phys. Rev.} C}

\def\PR{{\em Phys. Rev.}}

\usepackage{times}

\begin{document}
\vspace*{4cm}
\title{ Experimental constraints for additional short-range forces from neutron experiments}

\author{V. V. NESVIZHEVSKY}
\address{Institute Laue Langevin, 6, rue Jules Horowitz, Grenoble, France}

\author{G. PIGNOL, K. V. PROTASOV}
\address{LPSC, UJF - CNRS/IN2P3 - INPG, Grenoble, France}

\maketitle\abstracts{
We present preliminary results on sensitivity of experiments with slow neutrons to constrain additional forces in a wide distance range: 
from picometers to micrometers.
In the sub-nanometer range, available data on lengths of neutron scattering at nuclei provide the most competitive constraint.
We show that it can be improved significantly in a dedicated measurement of asymmetry of neutron scattering at noble gases. 
In the micrometer range, we present sensitivity of the future GRANIT experiment.
Further analysis will be presented in following publications.
}

\section{Introduction}

Our first analysis of constraints for additional fundamental short-range Yukawa-like forces 
\cite{Yao:2006px,Arkani-Hamed:1998rs,Antoniadis:1998ig,Rubakov:1983bb,Visser:1985qm,Antoniadis:1990ew,Lykken:1996fj} 
from neutron experiments was motivated by the measurement of gravitationally bound quantum states of neutrons 
\cite{Nesvizhevsky:2002ef,Nesvizhevsky:2003ww,Nesvizhevsky:2005ss}.
A conservative estimate \cite{Nesvizhevsky:2004qb} could be obtained from this experiment under the assumption that an additional force itself 
is not sufficiently strong for producing an additional quantum state near mirror surface, 
even in absence of the gravitational interaction.
Such constraint is valid in a very broad distance range from nanometers to micrometers 
and provides a kind of benchmark for short-range forces constraints from neutron experiments.
However, it is not as sensitive as the existing constraints obtained in experiments studying gravity at short distances or Casimir forces 
\cite{Kapner:2006si,Bordag:2001qi}.
On the other hand, the performed experiment with ultracold neutrons was not designed for searching for additional short-range forces 
and therefore dedicated neutron experiments could improve it and thus become rather competitive.

The natural condition for optimum sensitivity of neutron experiments 
is the approximate equality of the characteristic range of Yukawa-like force and the neutron wavelength.
This condition subdivides experiments with slow neutrons into 3 classes.

\begin{enumerate}
\item Scattering of cold (thermal) neutrons with the wavelength in the sub-nanometer range 
(the typical neutron wavelength is defined by the maximum in neutron spectra in moderators at nuclear reactors or spallation sources);
\item Neutron optics experiments with the neutron wavelength in the nanometer range 
(here the neutron wavelength is defined by typical values of effective Fermi-potential of $\approx 10^{-7}$ eV);
\item Gravitationally bound quantum states of neutrons with the largest available neutron wavelength (the characteristic distance range is $z_0=5.87 \ \mu$m).
\end{enumerate}

We will show that in each class of the experiments mentioned above one could improve constraints 
(relative to those in \cite{Nesvizhevsky:2004qb}) by many orders of magnitude in corresponding distance range.  
By coincidence, the third class of neutron experiments has optimum sensitivity to the characteristic distance range of micrometers, 
corresponding to predictions of theories with two extra spatial dimensions 
\cite{Yao:2006px,Arkani-Hamed:1998rs,Antoniadis:1998ig,Rubakov:1983bb,Visser:1985qm,Antoniadis:1990ew,Lykken:1996fj,Frank:2003ms}; 
and the first and second classes have optimum sensitivity in the nanometer range corresponding to theories with three extra spatial dimensions.

Neutron experiments provide a useful tool for studies of short-range forces as this has been noted in many publications 
\cite{Leeb:1992qf,Bertolami:2002xh,Abele:2003ga,Frank:2003ms,Watson:2004vh,Nesvizhevsky:2006,Greene:2006qj}.
Thus, neutron electric neutrality and slow velocity allow one to avoid systematic false effects related to electromagnetic forces. 
Several decades of methodical developments in this field resulted to extremely high precision of experiments.
For instance, one could note measurements of as small asymmetries of neutron scattering as $\approx10^{-8}-10^{-7}$ in ref \cite{Vesna:2005}.
Wavelengths of neutrons produced in nuclear reactors and spallation sources cover a very broad range of $~10^{-10}-10^{-5}$ m.
Finally, neutrons can be easily polarized to nearly 100 \% \cite{Kreuz},  
thus allowing studies of spin-dependent short-range interactions \cite{Baessler:2006vm}.
However, neutron methods are limited by important drawbacks.
Thus, one has to measure small absolute forces 
($\propto N$, where $N$ is the number of atoms in the test body at the distance $< \lambda$ where $\lambda$ is the characteristic range of the Yukawa-like force) 
in contrast to measurements of the interaction between two macroscopic bodies ($\propto N^2$), 
as for the Casimir-type experiments or for tests of gravity at short distances.
Neutron experiments are statistically limited. 
This is not so important a constraint for experiments with cold (thermal) neutrons with fluxes of up to 
$\approx 2 \times 10^{10}$ n/cm$^2$/s over cross-section of up to 6 cm by 20 cm \cite{Abele:2005xd}.
However, total fluxes of ultracold neutrons available for the experiments with gravitationally bound quantum states of neutrons are as low as $~10^{-2}$  s$^{-1}$. 
Neutron experiments are much ``smaller'' in terms of size and investments compared to high-energy physics experiments; on the other hand, 
these are not just ``table-top'' experiments.
Therefore, keeping in mind a rather few people involved and limited resources, one should accept that any progress takes time.

Let's analyze the present status of contraints for short-range forces from neutron experiments and possible prospects to improve their sensitivity. 
We will show that in the sub-nanometer range, available data on lengths of neutron scattering
at nuclei provide the most competitive constraint.
It can be improved significantly in a dedicated measurement of asymmetry of neutron scattering at noble gases. 
Neutron optics experiments will be analyzed in future publications. 
In the micrometer range we present sensitivity of the future GRANIT experiment. 
Fig.1 shows the existing neutron constraints and their possible improvements together with constraints known from other methods.

\begin{figure}
\begin{center}
\includegraphics[width=1\linewidth]{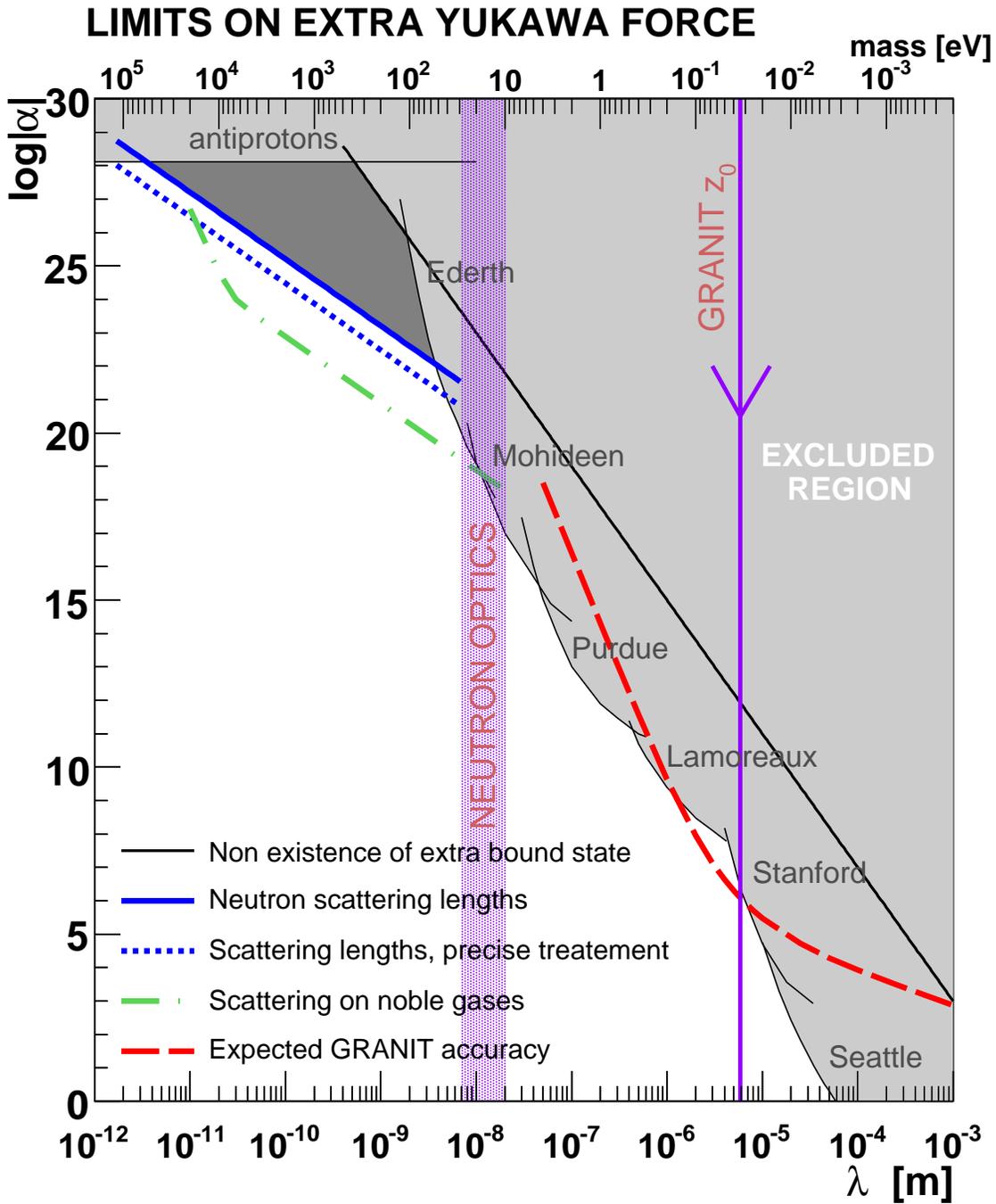}
\end{center}
\caption{
Constraints on Yukawa-like force from neutron experiments.
Strength of extra Yukawa-like force (normalized to gravitational interaction) 
is shown in function of the characteristic range of this force. 
Grey zone corresponds to the existing constraints obtained in measurements of gravity
at small distances, the Casimir force, 
the van der Waals force, in experiments with anti-protons. 
Dark zone indicates the constraints following from analysis of the neutron scattering lengths. 
The two vertical arrows show distance ranges for optimum sensitivity of the
neutron optics experiments (on left) and the experiments with gravitationally bound quantum states of neutrons (on right). 
Constraints from various neutron experiments are shown on the figure as follows: 
1) straight thin line indicates constraints obtained from the condition
of non-existence of an additional neutron quantum state near surface just due to the extra force alone;
2) dashed line shows possible constraints from GRANIT experiment; 
3) straight solid line corresponds to analysis of mass-dependence of neutron scattering
lengths, and dotted line shows possible improvements of this constraint; 
4) dashed-dotted line shows constraints which could be easily obtained in
measurements of asymmetry of neutron scattering at noble gases.
} \label{exclusion}
\end{figure}

\section{Sub-nanometer range: neutron scattering}

We will study effects of an additional Yukawa-like force $V_5(r)$ between a neutron (mass $m$) and a nucleus (mass $M = A m$). 
We assume that such a force is proportional to masses:
\begin{equation}
V_Y(r) = - G m M \alpha \frac{e^{-r/\lambda}}{r}
\end{equation}
 here $r$ is the distance, $G$ is the gravitational constant, $\alpha$ is the additional force strength normalized to gravity,
$\lambda$ is the range of the force.
The calculation of the scattering amplitude for a center of mass momentum $\hbar k$ is a standard exercise within the Born approximation:
\begin{equation}
\label{ScattAmplitude}
f_Y(\theta) = \frac{\alpha G m M}{\hbar^2} \frac{2 m \lambda^2}{1 + \left( 2 k \lambda \sin (\theta / 2) \right)^2}
\end{equation}
where $\theta$ is the scattering angle in the center of mass reference system.
As the range of a new interaction is much larger than nuclear range, 
the total amplitude for scattering due to nuclear interaction and new interaction is simply given by the sum 
$f(\theta) = f_N(\theta) + f_Y(\theta)$, where $f_N(\theta)$ is the nuclear term.

The main difficulty in extracting contraints from neutrons experiments consists in separating extra Yukawa-like force from the nuclear force. 
Such a difficulty is less important for other methods since they search for a deviation from known forces, 
such as the gravitational inverse square law, or the Casimir force.
The most safe way to do this separation is to use the fact that the nuclear force is very short-range.
The characteristic range of nuclear forces is as short as $\approx 1.2$ fm. 
However, they are much stronger than any additional short-range interaction of 
interest. 
Therefore we have to check that some "tails" of the nuclear force do not contribute 
to the additional Yukawa-like force analysis. 
We will use a simplified approximation of exponential decrease of nuclear 
forces as a function of distance; 
also we will neglect any "interference" between the two
forces. 
In this case, the contribution of nuclear forces will be negligible at distances $> 10^{-13}$ m.
In the low-energy limit 
(valid for slow neutrons discussed in the present paper) 
the neutron scattering at nuclei is described by a single 
parameter: 
the neutron scattering length $b$.
However, inspite of seeming simplicity of this interaction, 
one can not calculate precisely the values of scattering lengths 
"from first principles". 
Instead, the neutron scattering length is a phenomenological parameter, which
is measured experimentally for each particular isotop. 
Any additional force contribution can not be distinguished in general case from the nuclear force.
The method to distinguish these two interactions consists, 
for instance, in analysis of the wave-length dependence of neutron scattering.
The neutron-nuclear interaction is isotropic in the center-of-mass reference system, while an additional short-range force would favor forward
scattering compared to backward scattering, if
neutron wavelength is close to (or smaller than) 
the range of the additional force.
Another method consists in verification of presence of an additional linear term in the neutron scattering
length as a function of nuclear mass. Finally,
one could search for any force at distance significantly larger than the nuclear force range.

\subsection{Effect of the additional force on the scattering lengths}

Measurements of scattering lengths using interference methods\cite{Rauch} actually give access to the \emph{forward} scattering amplitude.
Therefore constraint for short-range forces can be obtained just from analysis of the mass dependence of the  neutron scattering length. 
Essentially, a contribution of an additional interaction (as a function
of nuclear mass) would result to an additional
linear term in the mass dependence of the neutron
scattering length,
\begin{equation}
b = f(0) = b_N + \frac{2 G m^2 M}{\hbar^2} \alpha \lambda^2.
\end{equation}
Let us now introduce explicit $A$ dependence of the scattering lenghts. 
In average, the nuclear scattering length for a nucleus of mass $m A$ is expected to be \cite{Peshkin}:
\begin{equation}
b_N(A) = R_0 A^{1/3}
\end{equation}
where the nuclear radius is known to be $R_0 \approx 1.2$ fm.
So, in presence of an additional force:
\begin{equation}
\label{scatteringA}
b(A) = R_0 A^{1/3} - \frac{2 m^3}{\hbar^2} G A \ \alpha \lambda^2
\end{equation}

\begin{figure}
\begin{center}
\includegraphics[width=0.7\linewidth]{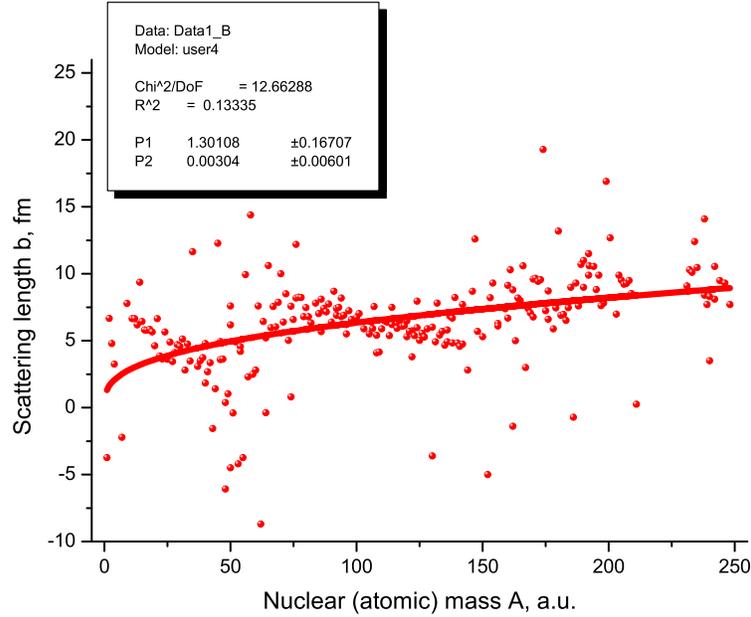}
\end{center}
\caption{
Scattering lenghts for neutron-nucleus interaction as a function of mass number. 
Fitted values for the parametrisation (\ref{scatteringA}) are presented for the parameters 
$R_0$ and $\frac{2 m^3}{\hbar^2} G \ \alpha \lambda^2$.} 
\label{Scatteringlengths}
\end{figure}

Fig. \ref{Scatteringlengths} shows the
known scattering lengths as a function of
atomic mass for various nuclei (isotops).
This function has been fitted according to the parametrization (\ref{scatteringA}). 
The nucleus radius $R_0$ is found to be $1.30 \pm 0.17$ fm, in accordance with the value known from nuclear physics.
Inspite of huge scattering of the scattering
length values, no significant linear contribution 
can be observed. We obtained the upper limit:
\begin{equation}
| \alpha | \ \left(\frac{\lambda}{1 {\rm nm}} \right)^2 > 1.6 \times 10^{23}
\end{equation}
This constraint is reported in fig. \ref{exclusion}. 
More sophysticated analysis taking into account nuclear shell structure can improve this limit by an order of magnitude, 
this will be achieved in a forthcoming publication.

\subsection{Asymmetry of neutron scattering at atoms}

Equation (\ref{ScattAmplitude}) suggests that if the neutron wavelenght is comparable to the additional force range, 
the scattering at a single nucleus would be anisotropic. 
A measurement of such forward-backward asymmetry would allow us to constrain additional forces with corresponding wavelength. 

The most promising method consists in a measurement of neutron scattering at diluted noble gases. 
In this case, however, one has to take into account that we deal with the scattering of neutrons at atomic electron cloud. 
This problematics is known in litterature as a measurement of neutron-electron scattering length $b_{ne}$ \cite{Fermi,Krohn}. 
The asymetry induced by the additional Yukawa-like force can be hidden by the asymetry due to neutron-electron interaction. 
Unfortunately, results of neutron experiments in the short-wavelength domain, aiming to measure the neutron-electron scattering
length $b_{ne}$, as well as their theoretical analysis are rather contradictory, 
and therefore they have been ignored by scientific community. 

That is why we have to be careful and conservative in order to obtain reliable constraints from previous experiments. 
So we decided to propose a neutron experiment free from systematic errors inherent for previous experiments. 
This is possible because 
the anisotropy due to neutron-electron interaction appears at the neutron wavelengths equal or smaller than atomic size
$\approx 10^{-11} {\rm m}$. 
So if the neutron wavelength is chosen to be larger than $\approx 10^{-11} {\rm m}$, the asymmetry will be free from neutron-electron contribution. 
Fortunately, this is just the domain of neutron wavelengths available at nuclear reactors and spallation sources. 

\begin{figure}
\begin{center}
\includegraphics[width=.7\linewidth]{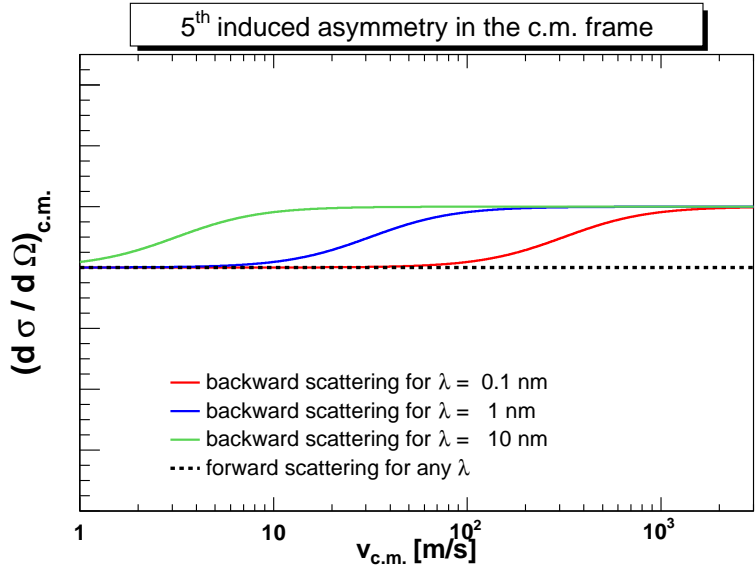}
\end{center}
\caption{
The asymmetry of neutron scattering at an atom in the center-of-mass reference system is shown 
in function of various characteristic ranges of the additional force $0,1; 1; 10$ nm.} 
\label{asymmetry}
\end{figure}

\begin{figure}
\begin{center}
\includegraphics[width=.42\linewidth]{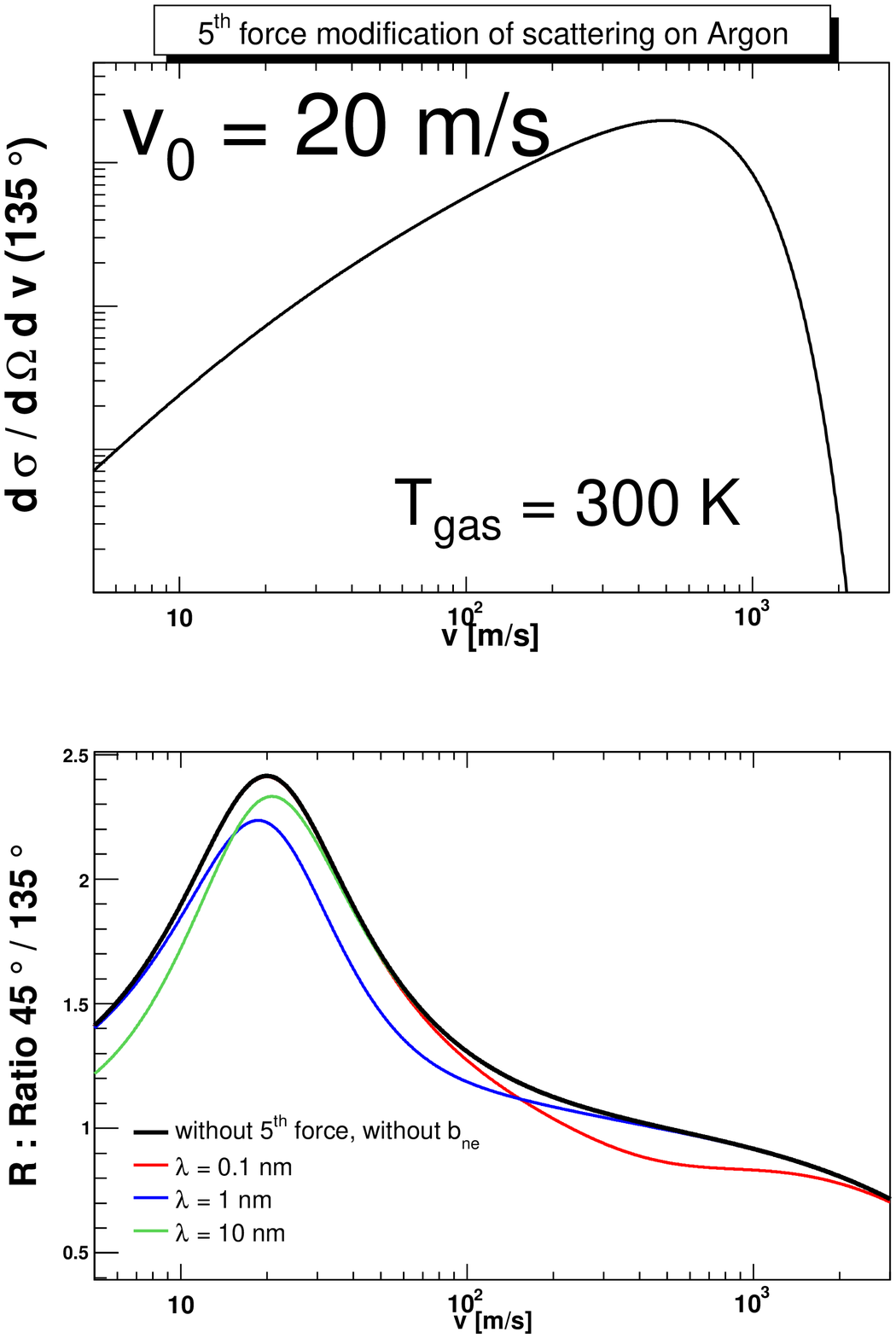}
\includegraphics[width=.42\linewidth]{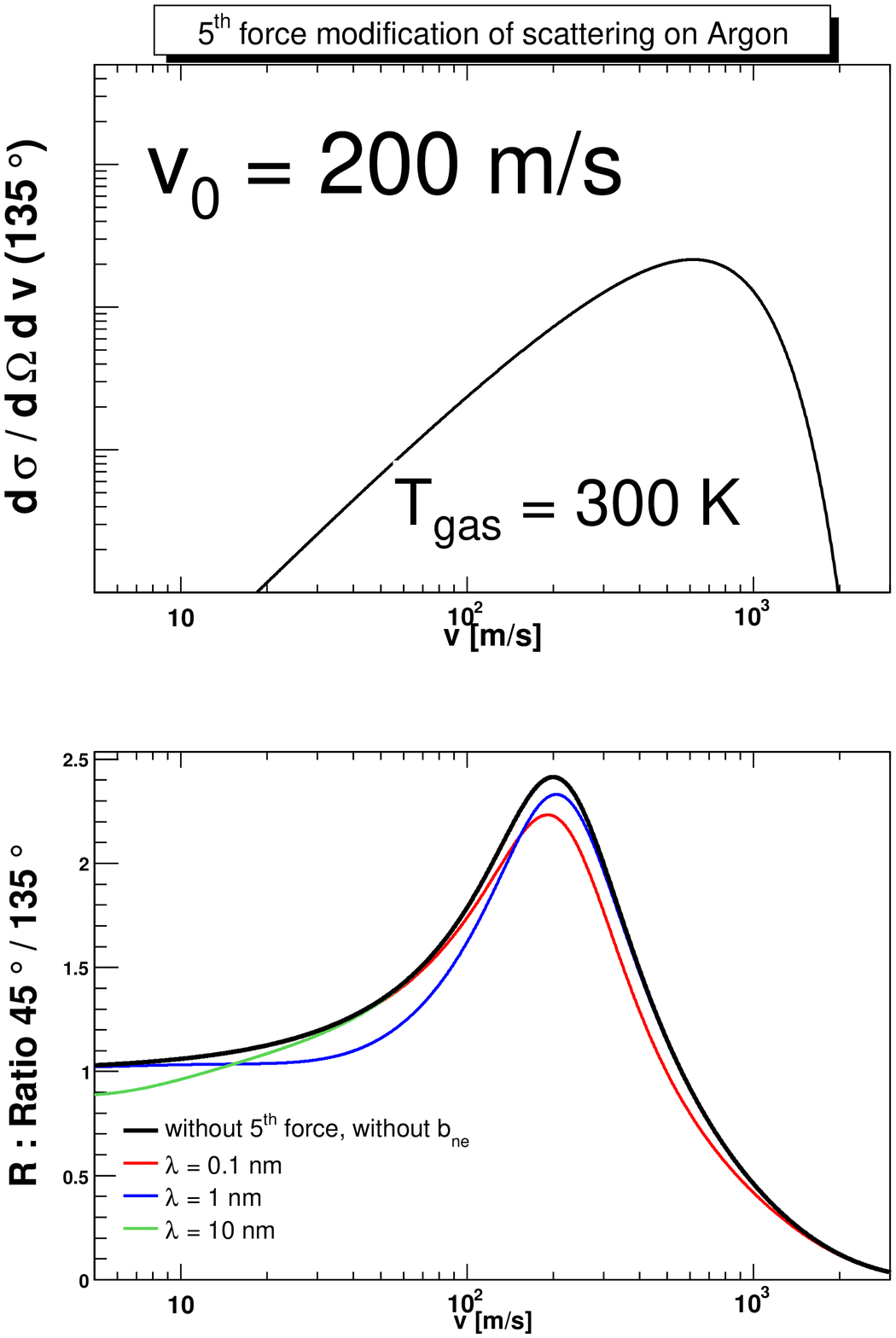}
\end{center}
\caption{
The ratio of neutron flux scattered at argon atom to 45 degree (forward) to that scattered to 135 degree (backward)
is shown on bottom figures in function of the final neutron velocity for the following cases: no additional forces, 
additional forces with the characteristic range of $0,1; 1; 10$ nm. 
The thermal motion of gas is taken into account. 
The two figures on top indicate the rate of collisions of neutrons 
with fixed initial neutron velocity with argon atoms (in thermal motion).
The initial neutron velocity for two pictures on left side is equal to $20$ m/s; 
the initial neutron velocity for two pictures on right side is equal to $200$ m/s.
} 
\label{thermal}
\end{figure}

Fig. \ref{asymmetry} shows the asymmetry of neutron scattering at an atom in the center-of-mass reference system 
in function of various characteristic ranges of the additional force. 
Evidently, one needs to use neutrons with velocities $1-1000$~m/s. 
This means that the thermal motion of the atoms can not be neglected, 
also that the neutron velocity in the laboratory fixed reference system is not equal to the neutron velocity in the center-of-mass reference system. 
Even totaly isotropic scatering in the center-of-mass reference system would be highly anisotropic in the laboratory system. 
Nevertheless, the kinematics of the scattering process is reconstructed precisely if one measures both the initial and the final neutron velocity 
in the laboratory reference system, 
that is easily feasible with reasonable statistical accuracy for thermal, cold, very cold and probably even ultracold neutrons. 
An advantageous geometry for the asymmetry measurement could be close to that used in 
\cite{Fermi,Krohn}; it allows to avoid major systematic errors. 
In this case, the calculated asymmetry would be modified by additional short-range forces as shown in fig.\ref{thermal} 
(the thermal motion of argon atoms is taken into account).

A measurement described above could easily provide accuracy of $10^{-3}$ for the ratio of forward to backward scattering probabilities; 
and a corresponding constraint for the additional short-range forces shown in fig. \ref{exclusion}.

\section{Micrometer range: gravitationally bound quantum states of neutrons}

A detailed anaysis of the limit on the additional short range force from the experiment which discovered neuton
bound states in the Earth's gravity field \cite{Nesvizhevsky:2002ef,Nesvizhevsky:2003ww} was presented in \ref{Nesvizhevsky:2004qb}.
The ``benchmark'' line in fig. \ref{exclusion} corresponds to the non existence of extra bound state of neutrons, 
additional to the bound states created by the Eearth gravity field and the bottom mirror.
Performing accurate measurement of the spectrum, the benchmark line can be improved in the micrometer range.
The aim of the GRANIT experiment is precisely to induce resonant transitions between bound quantum states, giving direct access to the spectrum. 
Besides statistics, the limiting factor for the accuracy of transition energies $\Delta E$ is the time $T$ during which harmonic perturbation is applied, 
according to Heisenberg's relation $\Delta E \ T > \hbar$.
The ultimate sensitivity of the GRANIT experiment corresponds to $T = 886$~s, that is, the $\beta$ decay lifetime of the neutron.
The corresponding precision of the transition frequencies is about $10^{-6}$.
To obtain the reach of the GRANIT experiment to contrain extra Yukawa-like force, we use perturbation theory to predict the shift of the $n$th level due to this force:
\begin{equation}
\delta E_n = 2 \pi G m \rho \alpha \lambda^2 \int_0^{\infty} dz |\psi_n(z)|^2 e^{-z/\lambda}(z)
\end{equation}
where $\psi_n(z)$ is the wave function of the $n$th level, and $\rho$ is the mass density of the bottom mirror.
The red dashed line in the exclusion plot fig. \ref{exclusion} corresponds to the constraint $( \delta E_2 - \delta E_1 )/E_1 < 10^{-6}$, 
assuming that the bottom mirror has five times the average mass density of the Earth.

\section{Conclusion}

We have presented preliminary results on sensitivity of experiments with slow neutrons to constrain additional forces in a wide distance range: 
from picometers to micrometers.
In the sub-nanometer range, available data on lengths of neutron scattering at nuclei provide the most competitive constraint and could be improved. 
Further increase in sensitivity could be achieved in a proposed dedicated measurement of asymmetry of neutron scattering at dilute noble gases; 
such an experiment would be free of a major possible systematic error caused by non-proper account for neutron-electron interaction.
In the micrometer range the sensitivity of the future GRANIT experiment could reach the accuracy of existing Casimir measurements if long times of neutrons 
in the quantum states will be achieved and the UCN density will be improved by a few orders of magnitude.
To summarize: 
experiments with slow neutrons could provide sensitive constraints for additional short-range forces and show strong potential for improvements. 
Further analysis will be presented in following publications.

\section*{Acknowledgments}

We are greteful to Agence Natonal de la Recherche, France, for supporting this project.

\end{document}